\documentclass[prl,twocolumn]{revtex4}
\usepackage{graphicx}
\usepackage{multirow}

\usepackage{amsmath}

\def\pket{|\Phi_a\rangle}
\def\ket#1{|#1\rangle}

\begin{document}
\title{Spin Hamiltonians with resonating-valence-bond ground states}
\author{Jennifer Cano}
\affiliation{Department of Physics, University of Virginia,
Charlottesville, VA 22904, USA}
\author{Paul Fendley}
\affiliation{Department of Physics, University of Virginia,
Charlottesville, VA 22904, USA}
\date{May 3, 2010}

\begin{abstract}
Quantum dimer models exhibit quantum critical points and liquid states
when the ground state is the resonating-valence bond (RVB) state. We
construct $SU(2)$-invariant spin-1/2 Hamiltonians with the same RVB
ground state. The main technical obstacle overcome is the fact that
different ``dimer'' configurations in the spin model are not
orthogonal to each other.  We show that the physics depends on how
dimers are related to the spins, and find a Hamiltonian that may be
quantum critical.
\end{abstract}

\maketitle

Quantum dimer models were designed to model antiferromagnets with a
strong tendency to form short-range spin-singlet valence bonds
\cite{rk}.  They can be obtained in certain large-$N$ limits from
antiferromagnets with $SU(N)$ or $Sp(N)$ symmetry \cite{RS}. An
important question is if the interesting physics found in quantum
dimer models -- e.g.\ quantum critical points with exactly computable
exponents, and resonating-valence-bond (RVB) liquids
\cite{rk,Moessner01} -- can be realized in spin-1/2 systems with an
unbroken $SU(2)$ symmetry.  A ``dimer'' in the spin model is a
nearest-neighbor singlet state, and a ``dimer configuration'' is a state where
each spin is paired into a singlet with one of its neighbors. The RVB
state is the equal-amplitude sum over all dimer configurations
\cite{Anderson}.


A crucial distinction between a quantum dimer model and the spin
system it models is that in the former, different dimer configurations
are orthogonal, whereas in the latter they are not. This difference
can dramatically change the physics \cite{LDA}. There have been a
variety of attempts to find quantum dimer physics in spin models (or
vice versa). One can decorate the lattice so that different dimer
states become more orthonormal as the amount of decorating is
increased \cite{RMS}. One also can define a spin model that resembles
the quantum dimer model, and study it numerically
\cite{Sandvik}. Another method is to expand around the orthogonal
limit, and develop a renormalization scheme to map the Heisenberg
model onto a generalized dimer model \cite{Schwandt}.

Here we take a different tack.  We construct a $SU(2)$-invariant spin
Hamiltonian with an exact RVB ground state. This is similar in spirit
to Refs.\ \onlinecite{fujimoto} and \onlinecite{seidel}, but differs
in detail.  Because of the non-orthogonality, there are two distinct
ways of defining spin models associated with a given quantum dimer
model. When the Hamiltonian includes a nearest-neighbor
antiferromagnetic Heisenberg term, the spin-spin correlation is
exponentially decaying \cite{LDA}. Changing the sign of this term
causes competing ferromagnetic and antiferromagnetic interactions,
likely resulting in very different physics.



We start by recalling some results for quantum dimer models with
Rokhsar-Kivelson (RK) Hamiltonians
\cite{rk,Moessner01}.  The Hilbert space in a quantum
dimer model is spanned by configurations of nearest-neighbor dimers on
some two-dimensional lattice, such that each site has a single dimer
touching it.  Each dimer configuration is orthogonal to all the
others.  Given a particular dimer configuration, a plaquette is
\textit{flippable} when each of its sites belongs to a dimer with an
adjacent site on the plaquette.  There are thus two possible ways a
plaquette $a$ can be flippable, which we denote as $|h_a\rangle$ and
$|v_a\rangle$. The RK Hamiltonian is
comprised of a sum of projectors, each of which acts only on the
dimers on a single plaquette. It includes a ``flip''
interchanging $|h_a\rangle$ and $|v_a\rangle$, the
simplest off-diagonal term possible in the quantum dimer model.
Letting $\pket \equiv
\ket{h_a}-\ket{v_a}$ and ${\cal P}^{\Phi}_a = \pket \langle {\Phi_a}|/2$ be
the projector onto this state, we have
\begin{equation}
\label{eq:HQDM}
H_{RK} = \sum_a {\cal P}^{\Phi}_a\ .
\end{equation}
Each term annihilates the sum $\chi_a\equiv\ket{h_a}+\ket{v_a}$, as
well as all states non-flippable on $a$.  Since each term in $H_{RK}$
is positive semi-definite, any state annihilated by all the terms is a
zero-energy ground state. The RVB state, the equal-amplitude sum over
all dimer states, is indeed such a ground state. The lattice and the
boundary conditions determine whether this ground state is unique.

One remarkable consequence of having a RVB ground state in a quantum
dimer model is that exact computations are possible, because
equal-time correlators in this ground state are the same as those in
the classical dimer model. The classical dimer model on any planar
lattice can be solved by a mapping onto free fermions
\cite{Kasteleyn,Fisher,fisher-stephenson}.  The resulting physics
depends on the lattice. The classical dimer model on a bipartite
lattice has a critical point where correlators decay
algebraically. The RK Hamiltonian at this point therefore must be
gapless and quantum critical \cite{Hastings}.  When the lattice is not
bipartite, the classical dimer model is not critical, and the quantum
model has topological order: the model is gapped, and the excitations
are fractionalized \cite{Moessner01}.

We turn to spin models, whose Hilbert space is comprised of a spin-1/2
particle at each site of a two-dimensional lattice. Our aim is to find
an $SU(2)$-invariant Hamiltonian with the RVB state as its ground
state.  By RVB state, we always mean the equal-amplitude sum over all
dimer configurations.  To define this uniquely in the spin model, we
must specify the convention for the sign of each dimer, i.e.\ which
spin state in the singlet has a plus sign in front of it, and which a
minus sign. For our purposes, we need to distinguish only between two
different kinds of RVB states, which we dub the positive-overlap and
negative-overlap RVB states, or PRVB and NRVB respectively. In the
PRVB state, the inner product between the two flippable configurations
on a single plaquette obeys $\langle h_a|v_a\rangle > 0$,
while in the NRVB state, $\langle h_a|v_a\rangle < 0$.

The
Hamiltonians we construct contain two types of terms: ``Klein'' terms
and ``flip'' terms.  Klein terms (at least attempt to) give an energy
to all non-dimer states \cite{klein}, whereas flip terms work
analogously to $H_{RK}$. All terms are written as products of various
$P^{s}({\Lambda})$, the $SU(2)$-invariant projector of the
spin-1/2 particles at sites in the set $\Lambda$ on to overall spin
$s$, i.e.\
$$P^{s}(\Lambda) = \prod_{j=r,j\ne s}^{|\Lambda|/2} 
\left(\vec{S}_{\Lambda} \cdot 
\vec{S}_{\Lambda} -
j(j+1)\right),
$$ where $\vec{S}_\Lambda=\sum_{\lambda \in \Lambda}\vec{S}_\lambda$
and $r=(1-(-1)^{2s})/4$.  Since a projector is positive semi-definite, a
zero-energy ground state is annihilated by each term individually.

A Klein term acts on spins at site $i$ and its nearest neighbors, a
set we denote as $N(i)$. In a dimer state, the spin $i$ and one of the
neighbors are in a singlet, so the spin of these $|N(i)|$
spins is less than $|N(i)|/2$.  The Klein Hamiltonian 
\begin{equation}\label{eq:HK} 
H_{K} = \sum_{i} P^{|N(i)|/2}({N(i)})
\end{equation}
then annihilates all dimer states, and sometimes certain non-dimer
states as well.  We say that the Klein term is
perfect if the null space of $H_K$ is exactly the dimer subspace.  It
was proved in \cite{cck} that on dimerizable subsets of the honeycomb
lattice, the square ladder, and the octagon-square lattice, the Klein
term is perfect with open boundary conditions.  (A dimerizable lattice
is defined as a lattice that can be completely covered in dimers.)  It
also appears to be perfect for subsets of the square lattice with open
boundary conditions.

Once a sign convention for the dimers is specified, the two flippable
plaquettes $|h_a\rangle$ and $|v_a\rangle$ are defined in the spin
model as in the dimer model, but these states are no longer
orthogonal.  While $\pket$ and $|\chi_a\rangle$ remain orthogonal,
$|\chi_a\rangle$ is no longer orthogonal to non-flippable plaquettes,
so ${\cal P}^{\Phi}_a$ no longer annihilates
it. Therefore using $H_{RK}$ in the spin model does {\em not} result
in the RVB state as a ground state. We thus must find a projection
operator that does not annihilate $\pket$, and that annihilates all
other dimer configurations on this plaquette, including
$|\chi_a\rangle$. It does not matter whether it annihilates non-dimer
configurations, as long as these are given an energy by the Klein
term.


We construct each flip term as a product $A_aF_a$ of projectors
obeying $[F_a,A_a]=0$. The projector $A_a$ annihilates all dimer
states on a plaquette $a$ that are not flippable, while
$F_a$ annihilates $|\chi_a\rangle$ but not $\pket$. Because $F_a$ and
$A_a$ commute, such a flip term satisfies the desired properties.
It is then straightforward to show that
the ground states of the Hamiltonian
\begin{equation}
H = H_K + \sum_a \left[F_aA_a + \overline{F}_a\overline{A}_a
\right]
\label{HKF}
\end{equation}
are the same as those of the quantum dimer model on the same lattice,
if the Klein term is perfect. We will show that we always can find
such pairs of projectors annihilating a NRVB state. For positive
overlap, we can sometimes find such pairs, and sometimes not.

\begin{figure}[t]
\begin{center}
\includegraphics[width=.2\textwidth,angle=0]{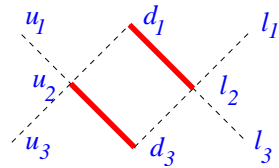}
\caption{The state $|v_a\rangle$ on the eight sites on which the
  modified flip term acts.}
\label{fig:8sites}
\end{center}
\end{figure}
Before presenting the general result, we discuss the square
lattice. The flip terms act on the eight spins illustrated in figure
\ref{fig:8sites}, the four spins comprising the plaquette $a$, and the
four nearest neighbors of two opposite corners. Any dimer
configuration must contain exactly two dimers on the links connecting
these eight spins, one dimer touching $u_2$ and the other $l_2$. If
the plaquette is flippable, then these two are on the plaquette; if
not, there must be at least one dimer on the links outside
$a$. Denoting by $U_a$ the set of sites ($u_1$, $u_2$, $u_3$), and
likewise for $L_a$ and $D_a$, the operator $A_a=P^{3/2}(U_a)
P^{3/2}(L_a)$ therefore annihilates any dimer configuration
non-flippable on $a$. (When the plaquette $a$ is at the boundary of
the lattice, this operator is modified to $P^{|U_a|/2}(U_a)
P^{|L_a|/2}(L_a)$.)  

To find $F_a$ for the square lattice, we exploit the fact that here,
$|\chi_a\rangle$ and $\pket$ are each proportional to either
the product of singlets or triplets on the {\em diagonals}. It is easy
to check by writing out the states explicitly that for positive
overlap $P^0(D_a)|\chi_a\rangle$=$0$ and
$P^0(D_a)\pket$=$\pket$, while for negative overlap
we have $P^1(D_a)|\chi_a\rangle$=$0$ and
$P^1(D_a)|\Phi_a\rangle$=$|\Phi_a\rangle$. Since $A_a$ does not act on
the spins $d_1$ and $d_3$, it commutes with both $P^0(D_a)$ and
$P^1(D_a)$. Thus $A_aP^0(D_a)$ annihilates PRVB states, and
$A_aP^1(D_a)$ annihilates NRVB states.  Thus for each overlap, we have
explicitly constructed a Hamiltonian of the form (\ref{HKF})
with a square-lattice RVB state as its ground state. The second term in
(\ref{HKF}) acts on the four spins of $a$ and the four spins
adjacent to the other two corners (i.e. the 90-degree rotation
of fig.\ \ref{fig:8sites}). 
Each term in
$H_K$ involves five spins, while each term $F_aA_a$ or
$\overline{F}_a\overline{A}_a$ involves 8 spins.
We note that the square-lattice
Hamiltonian proposed in Ref.\ \onlinecite{fujimoto} has
extraneous ground states, because it annihilates the state $\pket$ as
well as $|\chi_a\rangle$.


The projectors $A_a$ and ${\overline A}_a$ are easily generalized to
other lattices. The plaquette $a$ must be a polygon with an even
number $2n$ of sides, each connecting adjacent sites on the lattice.
For each site $i$ on the polygon, let $\alpha(i)$ be the set
consisting of $i$ and its nearest neighbors that are not adjacent on
the polygon. The operator $P^{|\alpha(i)|/2}(\alpha(i))$ then
annihilates any configuration with an ``external'' dimer connecting
$i$ to a site not adjacent on the polygon.  Labeling the sites on the
polygon consecutively, the operators
$$A_a= \prod_{i\ {\rm odd}} P^{|\alpha(i)|/2}(\alpha(i)),\quad
\overline{A}_a= \prod_{i\ {\rm even}} P^{|\alpha(i)|/2}(\alpha(i))
$$ each annihilate all dimer configurations not flippable on $a$.
Both work because all non-flippable dimer configurations on $a$ must
have the same number of external dimers touching odd sites on $a$ as
there are touching even sites. 

For negative overlap between the flippable plaquettes, flip operators
with the desired properties are
\begin{equation}
F_a=P^{n/2}(E_a),\qquad \overline{F}_a=P^{n/2}(O_a), 
\end{equation}
where $E_{a}$ and $O_{a}$ are the sets of the even sites and the odd
sites on the polygon respectively. Since $F_a$ acts only on the even
spins, it commutes with $A_a$, and likewise
$[\overline{F}_a,\overline{A}_a]=0$.  Let $|h_a\rangle$ be the
flippable plaquette where the dimers connect sites ($2j-1$, $2j$),
while $|v_a\rangle$ has dimers on sites ($2j-1$, $(2j-2)
\hbox{mod}\,n$). This determines these states up to an overall sign.
Up to this sign, the state $|h_a\rangle$ can be obtained from the
state $|v_a\rangle$ by shifting just the even spins by $2$ mod
$n$. Since $F_a$ acting on the $n$ even spins is the projector of
maximal spin, it produces a state completely symmetric in these
spins. We therefore have $F_a|h_a\rangle = \pm F_a|v_a\rangle$, with
the plus sign and minus signs occurring respectively for positive and
negative overlaps between $|h_a\rangle$ and $|v_a\rangle$.
Thus indeed $F_a|\chi_a\rangle=0$ for negative overlap.

We thus have found a Hamiltonian of the form (\ref{HKF}) with the NRVB
state as its ground state. This construction works for any lattice
where the sign convention can be chosen so that $\langle
h_a|v_a\rangle < 0$ for all $a$. The problem of constructing a
Hamiltonian with the PRVB state for an arbitrary lattice is still
open. Although our method works for the square lattice, it will not
work for lattices such as the honeycomb where the plaquette has $4j+2$
sites (such as the honeycomb): no $SU(2)$-invariant operator $F_a$
with the desired properties exists. However, in the honeycomb case,
the Hamiltonian of Ref.\ \onlinecite{fujimoto} works. Moreover, for
some non-bipartite lattices like the ``pentagonal'' lattice, there is
no way to have positive overlap for all plaquettes, only
negative \cite{RMS}.

One way to make a Klein term perfect is to ``decorate'' the lattice by
adding two spins on every link \cite{RMS}.  On such a decorated
lattice, the RVB state is annihilated by a Hamiltonian whose terms
involve only the spins on a given (decorated) plaquette $a$. As
before, the flipping operator must annihilate all dimer states save
the difference of the flippable configurations $|\chi_a\rangle$.
For an even number $2p$ of
sites on the undecorated lattice, we define
\begin{eqnarray*}
B_a &=& \prod_{j=1}^{p} (2-
P^0(a_{6j-2},a_{6j-1})-P^0(a_{6j+1},a_{6j+2})), \\ \overline{B}_a &=&
\prod_{j=1}^p (2- P^0(a_{6j+1},a_{6j+2})-P^0(a_{6j+4},a_{6j+5})),
\end{eqnarray*}
where the site labels $0,1,\dots 6p-1$ are interpreted mod $6p$. It is
straightforward to check that both $B_a$ and $\overline{B}_a$ indeed
annihilate all configurations not flippable on $a$, and that the
operators $F_a = B_a P^\Phi_a B_a$ and $\overline{B}_a =
\overline{B}_a P^\Phi_a \overline{B}_a$ have all the desired
properties. Thus the Hamiltonian $H= H_K + \sum_a
[F_a+\overline{F}_a]$ has the RVB state as a ground state.


We turn to the physics of the spin models with these RVB ground
states.  First, we note that the physics of the PRVB and NRVB states
is likely to be quite different. One reason we expect this is that
when $\langle h_a|v_a\rangle <0$, then the two N\'eel states of
alternating values of $S_z$ around the plaquette (i.e.\ $+-+-\dots$)
are absent from $|h_a\rangle + |v_a\rangle$. Thus one does not expect
an RVB state with this choice of overlap to be the ground state of a
Hamiltonian with purely antiferrogmagnetic interactions. This
assertion is straightforward to check by explicitly rewriting
Heisenberg Hamiltonians in the dimer basis \cite{RMS,Schwandt}; the
dimer flip term appears with the appropriate sign for the PRVB
state. Thus the physics of the NRVB state is likely to result from the
competition between ferrogmagnetic interactions and the
antiferromagnetic interactions from the Klein terms and the other
interactions between spins farther than nearest-neighbor.

Many properties of the PRVB state have already been derived
\cite{LDA,Shapir}, because all diagonal equal-time correlators can be
written as correlators in a {\em classical} two-dimensional model.
For the quantum dimer model, the model is simply the classical dimer
model. This follows from the fact that different dimer configurations
are orthogonal. Letting ${\cal O}({\cal D})$ be the value of the
operator ${\cal O}$ in the dimer configuration ${\cal D}$, we have
$$
\langle {\rm RVB} | {\cal O} | {\rm RVB}\rangle 
= N_D^{-1}{\sum_{{\cal D}} {\cal O}({\cal D})}{}\ ,
$$
where $N_D$ is the number of different dimer 
configurations, the partition function of the classical dimer model.

To generalize this to the spin model, we need to take into account the
non-orthogonality of the inner product in the dimer basis. The inner
product of two dimer configurations ${\cal D}$, ${\cal D}'$ has a 
nice geometric description \cite{Sutherland}.  If we place both
configurations on the same lattice, we then can form loops by gluing
the dimers from ${\cal D}$ and ${\cal D}'$ together at each site.
These are dense loops with exactly one loop touching each site.
Then $\langle {\cal D}|{\cal D}'\rangle = \pm 2^{n_{\cal L}+n_{2}-
n_{\cal D}}$, where $n_2$ is the number of loops of length 2 formed by
this gluing (i.e.\ the number of links covered by a dimer in both
${\cal D}$ and ${\cal D}'$), $n_{\cal L}$ is the number of closed
loops of length greater than 2, and $n_{\cal D}$ is the number of
dimers \cite{Sutherland}. The $2^{-n_d}$ arises from the normalization
of the singlet state, while the factor of $2$ for each loop arises
because there are two non-vanishing terms in the inner product coming
from alternating values of $S_z$ around the loop. Thus in the spin
model \cite{Shapir},
\begin{equation}
\langle {\rm PRVB} | {\cal O} | {\rm PRVB}\rangle 
= \frac{\sum_{{\cal L}} 2^{2n_{\cal L}+n_2}{\cal O}({\cal L})}
{\sum_{\cal L} 2^{2n_{\cal L}+n_2}}\ ,
\label{loopPRVB}
\end{equation}
where each sum is over all closed loop configurations ${\cal L}$
formed by gluing two dimer configurations together. The extra factor
of 2 for loops of length longer than 2 arises because there are two
dimer configurations corresponding to each such loop.  Equal-time
correlators in the spin model with PRVB ground state thus identical to
those in a classical dense loop model with weight $4$ per loop (the
effect of the length-2 loops can be absorbed in a weight per unit
length of the longer loops \cite{Schwandt}).  The only terms
contributing to the spin-spin correlator are those where both spins
are on the same loop: when the two spins are on different loops, the
sum over the two alternating-$S_z$ values on each loop makes this
contribution vanish.

Classical loop models, often known as the $O(N)$ loop models for
weight per loop $N$, have been studied extensively. When the loops are
on a two-dimensional bipartite lattice, the model can be mapped onto a
Coulomb gas, and when $|N|\le 2$, it has a critical phase
\cite{Nienhuis}. For $N>2$, correlators decay exponentially:
heuristically, the energy from creating more loops always
beats the entropy gain from creating long loops. For the PRVB state,
$N$=$4$, and numerics for the square lattice indicate that the
spin-spin correlator in the PRVB state indeed decays exponentially
with a small correlation length \cite{LDA}. This seems very different
from the quantum dimer model, which on the square lattice with RK
Hamiltonian is critical \cite{rk}. However, one can gradually go from
spins to dimers by using decorated lattices \cite{RMS}, and there is
no particular indication that the physics changes dramatically as the
amount of decoration is increased. One way of reconciling these observations
is if some quantities (e.g.\ dimer-dimer correlations) behave
as they do in the dimer model, even though the spin-spin correlations
decay exponentially.

We expect that the physics of the PRVB and NRVB states is very
different, even though they correspond to the same quantum dimer
model. Different inner products results in different physics for
quantum dimer models on the Kagome lattice \cite{Misguich2}, and for
quantum loop models \cite{PFnets}.  For the NRVB state, the expression
for correlators is the same as (\ref{loopPRVB}), except for important
relative signs. For the square lattice with open boundary conditions,
each plaquette flip changes the total number of loops by $\pm
1$. Therefore on the square lattice, each loop gets a {\em negative}
weight:
\begin{equation}
\langle {\rm NRVB} | {\cal O} | {\rm NRVB}\rangle 
= \frac{\sum_{{\cal L}} (-4)^{n_{\cal L}}(-2)^{n_2}{\cal O}({\cal L})}
{\sum_{\cal L} (-4)^{n_{\cal L}}(-2)^{n_2}}\ .
\label{loopNRVB}
\end{equation}
Negative weights can cause cancellations between different
configurations. For example, as we saw, the N\'eel state around a
plaquette does not appear in $|\chi_a\rangle$ with negative overlap.
Classical loop models with $N<-2$ do not seem to have been studied. Given
that the classical dimer model is critical, and that size of the critical
phase of the loop model gets larger as $N$ is decreased \cite{Nienhuis}, it
is possible that even spin-spin correlators in the NRVB state
will be critical. Thus even though the NRVB state does not arise from
purely antiferromagnetic interactions, the possibility that it could
yield an $SU(2)$-invariant quantum critical point makes it a worthy
subject for further study.

\bigskip

This work was initiated while we were at the Rudolf Peierls Centre for
Theoretical Physics, University of Oxford; we are grateful for the
hospitality. We thank Roderich Moessner and Shivaji Sondhi for
numerous conversations, and Fabien Alet and Nick Read for closing a
major gap in our knowledge.  This work was supported by
a Harrison award from the University of Virginia (J.C.), and by the
NSF grant DMR/MSPA-0704666 (P.F.).



\end{document}